\documentclass[%
reprint,
superscriptaddress,
amsmath,amssymb,
aps,
pra,
 floatfix
]{revtex4-2}

\usepackage{graphicx}        
\usepackage{dcolumn}         
\usepackage{bm}               
\usepackage{hyperref}   
\usepackage[mathlines]{lineno}
\usepackage{xcolor} 
\usepackage[capitalize]{cleveref} 
\usepackage{ragged2e}
\usepackage{microtype}     
\usepackage{amsfonts,mathrsfs} 
\usepackage{booktabs}          
\usepackage{tikz}            
\usepackage{comment}          
\usepackage{subfig}
\usepackage{ragged2e}

\begin{document}

\title{Statistical mechanics of multipartite entanglement in hypergraph states
}

\author{Paolo Scarafile}
\affiliation{Dipartimento di Fisica, Università di Napoli, I-80126 Napoli, Italy}
\affiliation{INFN, Sezione di Bari, I-70125 Bari, Italy}

\author{Giorgia Trotta}
\affiliation{Dipartimento di Fisica, Università di Napoli, I-80126 Napoli, Italy}
\affiliation{INFN, Sezione di Bari, I-70125 Bari, Italy}

\author{Paolo Facchi}
\affiliation{INFN, Sezione di Bari, I-70125 Bari, Italy}
\affiliation{Dipartimento di Fisica, Università di Bari, I-70126 Bari, Italy}

\author{Giuseppe Magnifico}
\affiliation{INFN, Sezione di Bari, I-70125 Bari, Italy}
\affiliation{Dipartimento di Fisica, Università di Bari, I-70126 Bari, Italy}

\author{Giorgio Parisi}
\affiliation{Dipartimento di Fisica, Università degli Studi di Roma La Sapienza}
\affiliation{Istituto Nazionale di Fisica Nucleare, Sezione di Roma I}
\affiliation{Institute of Nanotechnology (NANOTEC) - CNR, Rome unit}

\author{Saverio Pascazio}
\affiliation{INFN, Sezione di Bari, I-70125 Bari, Italy}
\affiliation{Dipartimento di Fisica, Università di Bari, I-70126 Bari, Italy}

\author{ Karol Życzkowski}
\affiliation{Institute of Theoretical Physics, Faculty of Physics, Astronomy and Applied Computer Science, Jagiellonian University
Ul. Lojasiewicza 11, 30-348 Kraków, Poland}
\affiliation{Center for Theoretical Physics (CFT), Polish Academy of Sciences,
 Warszawa, Poland}

\begin{abstract}

We investigate multipartite entanglement in a particular family of pure $n$-qubit hypergraph states
through a statistical-mechanics framework, where the average bipartite purity maps onto an effective Hamiltonian of $2^n$ classical binary  spins. In this correspondence, each hypergraph state uniquely corresponds to a classical spin configuration, while temperature serves as a control parameter that continuously interpolates between a uniform ensemble of random hypergraph states at high temperature and maximally multipartite entangled states (MMES) at zero temperature. Remarkably, the exponential of the zero-temperature entropy directly gives the number of MMES within the set of hypergraph states. For small system sizes ($n \leq 5$), we perform an exact enumeration, fully characterizing the energy landscape and associated thermodynamic observables, and validating known MMES counts. For larger systems ($n = 6$ and $7$), where exact methods become computationally infeasible, we employ simulated annealing and parallel tempering algorithms to efficiently sample the exponentially large state space. Our analysis yields quantitative predictions of the number of MMES  and reveals how entanglement is statistically distributed across the sets of hypergraph states. These results establish hypergraph states as an ideal platform for investigating multipartite entanglement through thermodynamic methods, offering both computational advances and physical insights into the structure of quantum entanglement in restricted families of quantum states.
\end{abstract}

\maketitle

\section{Introduction}\label{sec:introduction}
Quantum entanglement represents one of the most distinctive phenomena in quantum physics, serving as an essential ingredient for numerous quantum information protocols \cite{Bruss_2002, Horodecki_2009}.
Its applications span a wide range of emerging quantum technologies, from cryptographic schemes and quantum communication systems to quantum computers and simulators \cite{Ekert_1991, Bennett_1993, Ekert_1996,  Fuchs_1997, Bouwmeester_1997,  Boschi_1998, Nielsen_2000, Jozsa_2003,  Horodecki_2001, Bennett_2014, Zhao_2025}. 
The theoretical framework for characterizing entanglement varies significantly depending on the number of parties making up the system.

For pure quantum states under bipartition, the von Neumann entropy of the reduced density matrix provides a complete and well-established characterization and quantification method  \cite{Bennet_1996, Wootters_1998, Wootters_2001, Amico_2008}. 
However, extending this understanding to multipartite entanglement presents considerable theoretical challenges, as no single unified framework has emerged \cite{Coffman_2000, Wong_2001, Meyer_2002,  Jakob_2007, Facchi_Max_multi_ent_2008, Amico_2008}. 
Furthermore, understanding the structure of multipartite entanglement in many-body systems remains challenging due to the exponential growth of the Hilbert space dimension with system size, and the presence of intriguing phenomena, such as the emergence of frustration, i.e. the impossibility of simultaneously achieving maximally entanglement across all bipartitions of the system \cite{Rains_1999, Higuchi_2000, Scott_2003, Scott_2004, Facchi_frustartion_2010, Huber_2017, Steinberg_2024}. 

Given this complexity, statistical approaches involving the distribution of entanglement measures, particularly the purity over all balanced bipartitions of the system, provide a crucial framework for analyzing multipartite entanglement \cite{Facchi_Max_multi_ent_2008, Facchi_Probability-density-function_2006, Facchi_Charac_measuring_mult_ent_2007}, as shown in the context of random pure states sampled according to the Haar distribution \cite{Lubkin_1978, Lloyd_1988, Page_1993, Zyczkowski_2001, Scott_2003, Giraud_2007, Facchi_Statist_mechanics_2009, Facchi_Lincei_2009, Facchi_Classical-statistical-mechanics_2010}.

The properties of multipartite entanglement are particularly relevant when searching for an important class of states, the maximally multipartite entangled states (MMES)\cite{ Facchi_Max_multi_ent_2008}, i.e., those states that maximize the entanglement distribution across all balanced bipartitions
\cite{Facchi_Lincei_2009, Facchi_Local_Ham_for_mmes_2010, Gonzalez_2012, DeVicente_2013, Zyczkowsi_2015, Zyczkowsi_2018, Burchardt_2020,  Zyczkowsi_2021, Zyczkowsi_2022, Bernards_2022, Zyczkowski_2023}. 

Identifying and characterizing MMES is a formidable task, both analytically and numerically, due to the exponentially large number of parameters and bipartitions to handle as a function of the system size. 
A promising strategy to address these challenges is to focus on structured ensembles of quantum states that preserve non-trivial entanglement properties and display symmetries and constraints suitable for systematic analysis. 

Uniform states, characterized by equal weights in the computational basis, turn out to be compatible with MMES formal conditions for qubit systems \cite{Facchi_Lincei_2009, Goyeneche_2014}. Thus, these states represent a promising ensemble for investigating multipartite entanglement.

In this work, we study a specific set of uniform states, namely the class of  states whose amplitudes in the computational basis are restricted to $\pm 1$, up to normalization. It is known that these states are in a one-to-one correspondence with hypergraph states \cite{Macchiavello_2011, Qu_2013, Macchiavello_2013, Macchiavello_2014,  Hamma_2022, Macchiavello_2026}. These  states can be generated using multi-controlled Pauli-z operations, experimental produced by \cite{Huang_2024} and represented by mathematical hypergraphs, in which each vertex corresponds to a qubit in the system and the hyperedge to a multi-controlled Z operator.

These states form a discrete set embedded within the Hilbert space and provide a natural setting for statistical analysis. Moreover, as recently highlighted in \cite{Trotta_2026}, they show an enhanced tendency of sampling highly entangled configurations.

We consider a statistical mechanics framework \cite{Facchi_Statist_mechanics_2009} for analyzing multipartite entanglement by mapping the average bipartite purity, our measure of multipartite entanglement, onto an effective Hamiltonian of a classical spin system with binary variables \cite{Mezard_1986, Mezard_1986.2, Mezard_1987, Mezard_2001, Parisi_2024}. This mapping transforms the quantum entanglement problem into a classical statistical mechanics one, where each hypergraph state corresponds to a spin configuration, and the entanglement structure emerges from four-body spin interactions. 

By introducing a  temperature parameter, we study thermodynamic observables such as internal energy and entropy to characterize the entanglement landscape. Remarkably, in this framework, the zero-temperature entropy provides a direct count of the number of MMES within the set of hypergraph states, while finite-temperature behavior reveals the statistical distribution of multipartite entanglement. Thus, inverse temperature $\beta$ acts as a tunable parameter: $\beta=0$ corresponds to 
a uniform ensemble of random hypergraph states, whereas the large positive and negative values of $\beta$ select maximally multipartite entangled states and fully separable states, respectively.

For small systems with $n \leq 5$, we exhaustively enumerate all hypergraph states, allowing a complete characterization of the state space and a precise computation of thermodynamic quantities. For larger systems with $n = 6$ and $7$, exact enumeration becomes infeasible (see Table ~\ref{tab:sub-sets}), so we perform extensive numerical simulations employing Simulated Annealing (SA) \cite{metropolis_1953, Annealing_1983, cerny_1985, Laarhoven_1987, Ingber_1993} and Parallel Tempering (PT) \cite{Tempering_1986, Geyer_1991, Marinari/Parisi_1992, Tempering_1996, Falcioni_1999, Grigera_2001, Earl_2005}  algorithms. These methods efficiently sample the relevant statistical ensembles and enable us to estimate thermodynamic observables across a broad range of temperatures  

Our results demonstrate that the statistical
mechanics framework captures key features of multipartite entanglement of hypergraph states, providing both a conceptual foundation and practical computational tools to investigate quantum complexity and entanglement in structured subspaces through classical thermodynamic analogies.

The paper is structured as follows. In Sec.~\ref{sec:sec1}, we establish the quantum-classical mapping and derive the effective Hamiltonian of the classical statistical mechanics model defined over the hypergraph-state discrete space. In Sec.~\ref{sec:sec2}, we present exact computations of thermodynamic observables for small systems. In Sec.~\ref{sec:sec3}, we introduce the Parallel Tempering (PT) algorithm for larger system sizes. In Sec.~\ref{sec:sec4}, we discuss the validation procedures of the  results. Finally, in Sec.~\ref{sec:disc} and Sec.~\ref{sec:conclusion}, we draw our conclusions and outline future perspectives.

\section{From qubits to classical spins}\label{sec:sec1}

An $n$-qubit hypergraph state is represented by a normalized vector in the Hilbert space $\mathcal{H} = \mathbb{C}^{2^{n}}$, whose computational-basis amplitudes have equal magnitude and, up to an overall phase, signs in the discrete set $\{+1, -1\}$~\cite{Macchiavello_2011,  Macchiavello_2026}.

Thus, it can be written as
\begin{equation}
|\psi(\bm{s})\rangle = \frac{1}{2^{n/2}} \sum_{k \in \mathbb{B}^n} s_k |k\rangle, \qquad s_k \in \{+1, -1\},
\label{eq:unifstate}
\end{equation}
where $\mathbb{B}=\{0,1\}$ and the vector $\bm{s} = (s_k)_{k\in\mathbb{B}^n}$ encodes the sign configuration over the  $2^n$ computational basis states $|k\rangle$. Measurements in the computational basis therefore yield a uniform probability distribution over all possible outcomes. Hypergraph states also exhibit distinctive structural and entanglement properties, as recently pointed out in~\cite{Trotta_2026}. The total number of physically distinct hypergraph states is:
\begin{equation}
    N_{\mathrm{hyper}} = 2^{2^n -1}.
    \label{eq:N_hyper}
\end{equation}
The corresponding values for different system sizes $n$ are reported in Table~\ref{tab:sub-sets}.
\begin{table}[t]
\begin{ruledtabular}
\begin{tabular}{cc}
\textbf{Number of Qubits ($n$)} & \textbf{Number of States} \\
\hline
        3  &  128  \\
        4  &  32{,}768  \\
        5 & $2{,}147{,}483{,}648$  \\
        6  & $9.22\times 10^{18}$  \\
        7  &  $1.70 \times 10^{38}$  \\
        8  & $5.79 \times 10^{76}$  \\
        9  & $6.70 \times 10^{153}$  \\
        10  & $8.99 \times 10^{307}$  \\
        11  & $ 1.62 \times 10^{616}$ \\
\end{tabular}
\end{ruledtabular}
\caption{\justifying Cardinality of the set of hypergraph states grows double exponentially as a function of the number of qubits $n$.}
\label{tab:sub-sets}
\end{table}

To investigate the structure of multipartite entanglement within this set, we employ the purity of reduced density matrices. For any bipartition of the $n$-qubits into subsystems $A$ and $\bar{A}$, given $\rho=| \psi \rangle \langle \psi |$, the purity $\pi_{A}(|\psi\rangle) = \mathrm{Tr} \left ( \rho^2_{A} \right )$, where $\rho_A = \mathrm{Tr}_{\bar{A}} \left ( \rho \right)$, quantifies the degree of entanglement between the two subsystems. The purity takes values in the interval $1/2^{n_A} \leq \pi_A \leq 1$, where 
$n_A=|A|$ is the number of qubits in $A$. Lower values of $\pi_A$ indicate
stronger entanglement, where the situation of maximum entanglement between the two partitions $A$ and $\bar{A}$ is represented by the minimum value of $\pi_A = 1 / 2^{n_A}$. 
Let us recall that this quantity describes the entanglement of a single bipartition. For a global description of multipartite entanglement for pure states, it is possible to introduce the \emph{potential of multipartite entanglement} which averages this quantity over all balanced bipartitions (i.e., such that $n_A = \lfloor n/2 \rfloor $) of the $n$-qubits systems \cite{Facchi_Lincei_2009, Facchi_Max_multi_ent_2008}: 

\begin{equation}
\pi_{\textrm{ME}}(|\psi\rangle) = \binom{n}{\lfloor n/2 \rfloor}^{-1} \sum_{|A|= \lfloor n/2 \rfloor} \pi_{A}(|\psi\rangle)
\label{eq: multpotentialdef}
\end{equation}

For an $n$-qubit hypergraph state~\eqref{eq:unifstate} , this quantity admits the elegant representation:
\begin{equation}
\pi_{\mathrm{ME}}(\bm{s}) = \frac{1}{2^{2n}}\!\! \sum_{k,{k'},l,{l'}\in\mathbb{B}^n}\!\!\! \Delta(l, l'; k, k')\, s_l s_{l'} s_k s_{k'},
\label{eq:potential-unif}
\end{equation}
where \(\Delta(l, l'; k, k') = \Delta(l, l'; k, k'; \lfloor n/2 \rfloor)\) is a coupling function depending on the number of qubits $n$,
\begin{align}
    \Delta(l, l'; k, k')
=\frac{1}{2}\tilde{\Delta}(l, l'; k, k')  
 + \frac{1}{2}\tilde{\Delta}(l', l; k, k' ),
\end{align}
where
\begin{align}
    \tilde{\Delta}(l, l'; k, k') &= \binom{n}{\lfloor \frac{n}{2} \rfloor}^{-1} \!\!\sum_{|A|= \lfloor n/2 \rfloor}\!\! \delta_{l_{A} k'_{A}} \delta_{l'_{A} k_{A}} \delta_{l_{\bar{A}}k_{\bar{A}}} \delta_{l'_{\bar{A}}k'_{\bar{A}}}.    
\end{align}

Within this framework, maximally multipartite entangled states (MMES) correspond to configurations that minimize the potential $\pi_{\mathrm{ME}}$. However, identifying these states remains challenging due to the exponential scaling of both the number of bipartitions and the configuration space with system size.

By making use of the one-to-one correspondence between  hypergraph states  and  sign vectors $ \bm{s} = (s_k)_{k\in\mathbb{B}^n}$, the problem of multipartite entanglement can be recast in terms of classical variables. Each configuration \( \bm{s} \) can be interpreted as a configuration of a classical model with \( 2^n \) effective binary spins. This reformulation enables the application of statistical mechanics tools to the study of multipartite quantum correlations. We formally define the effective classical Hamiltonian~\cite{Facchi_Statist_mechanics_2009} as
\begin{equation}
H(\bm{s}) = \pi_{\mathrm{ME}}(\bm{s}),
\label{eq:hamiltonian}
\end{equation}
where the coupling function \(\Delta(l, l'; k, k')\) acts as a  symmetric interaction kernel encoding four-body correlations. Unlike conventional Ising models with two-body interactions, this Hamiltonian exhibits long-range four-spin couplings, reflecting the genuinely multipartite nature of quantum entanglement. 

To probe the statistical properties of the classical system, extending the approach of Refs.~\cite{Facchi_Statist_mechanics_2009,Facchi_Classical-statistical-mechanics_2010} to the study of hypergraph states, we introduce a canonical ensemble with inverse temperature $\beta = 1/T$. 
The associated partition function is defined as
\begin{equation}
\mathcal{Z}(\beta) = \sum_{\bm{s}}e^{-\beta H(\bm{s})} =\int dE  P(E) e^{-\beta E},
\label{eq:partition-function}
\end{equation}
and is the Laplace transform of the density of states,
\begin{equation}
P(E) = \sum_{\bm{s}} \delta\left(H(\bm{s}) - E\right),
\label{eq:statedens}
\end{equation}
which enumerates the number of binary spin configurations $\bm{s}$, and hence hypergraph states, at energy $E$ (that is with potential of multipartite entanglement equal to~$E$).

Thermodynamic observables, such as the internal energy $U(\beta)$ and entropy $S(\beta)$, can be systematically derived from the partition function $\mathcal{Z}(\beta)$. In particular, the internal energy corresponds to the ensemble average of the Hamiltonian and is given by:
\begin{align}
    U(\beta) &= \langle H(\bm{s})\rangle_{\beta}  
    \nonumber\\
    &= \frac{1}{\mathcal{Z}(\beta)} \int dE P(E) E e^{-\beta E} = -\frac{\partial}{\partial \beta} \ln  \mathcal{Z}(\beta). 
\label{eq:internal-energy}      
\end{align}
This expression yields the thermal average of the entanglement potential at inverse temperature $\beta$. 

The Helmholtz free energy is defined in the canonical ensemble as
\begin{equation}
F(\beta) = -\frac{1}{\beta} \ln \mathcal{Z}(\beta),
\end{equation}
from which the entropy follows via the thermodynamic relation
\begin{align}
S(\beta) = \beta \bigl(U(\beta) - F(\beta)\bigr) 
                  = \beta\, U(\beta) + \ln \mathcal{Z}(\beta).
\label{eq:entropy1}
\end{align}
Alternatively, by differentiating~\eqref{eq:entropy1}, and using~\eqref{eq:internal-energy},  we get the differential relation
\begin{equation}
\frac{\partial S}{\partial \beta} = \beta \frac{\partial U}{\partial\beta},
\end{equation}

which can be integrated to get
\begin{align}
S(\beta) = S_0 + \int_0^\beta \beta' \frac{\partial U}{\partial \beta'} d\beta' = S_0 + \beta U(\beta) - \int_0^\beta d\beta' U(\beta') .
\label{eq:entropy2}
\end{align}
In particular, the entropy at $\beta=0$ is given by $S_0 = \ln \mathcal{Z}(0)$, corresponding to the logarithm of the total number of accessible microstates, i.e., the cardinality of the full set of hypergraph states. At finite $\beta$, the entropy $S(\beta)$ quantifies the effective number of quantum states thermally accessible at that temperature:
\begin{equation}
\Omega(\beta) = e^{S(\beta)}.
\label{eq:number-ofstate}
\end{equation}
In the low-temperature regime, $\beta \to +\infty$ ( i.e., $T \to 0^+$), $\Omega(\beta)$ provides the  number of hypergraph MMES of $n$ qubits.

\section{Exact Computation of Thermodynamics for Small Systems}\label{sec:sec2}

\begin{figure*}[t]
    \centering

    \subfloat[\label{fig:hist3}][]{%
        \includegraphics[width=0.3\textwidth]{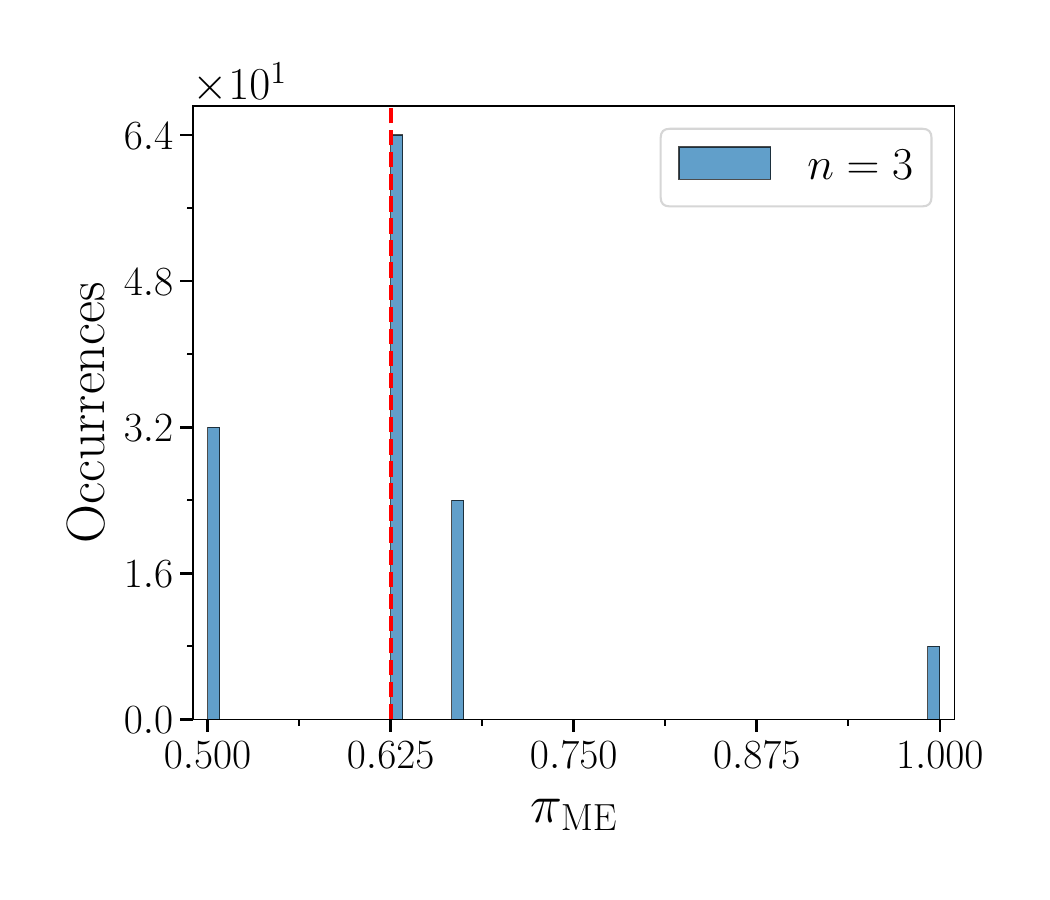}%
    }\hspace{0.03\textwidth}
    \subfloat[\label{fig:hist4}][]{%
        \includegraphics[width=0.3\textwidth]{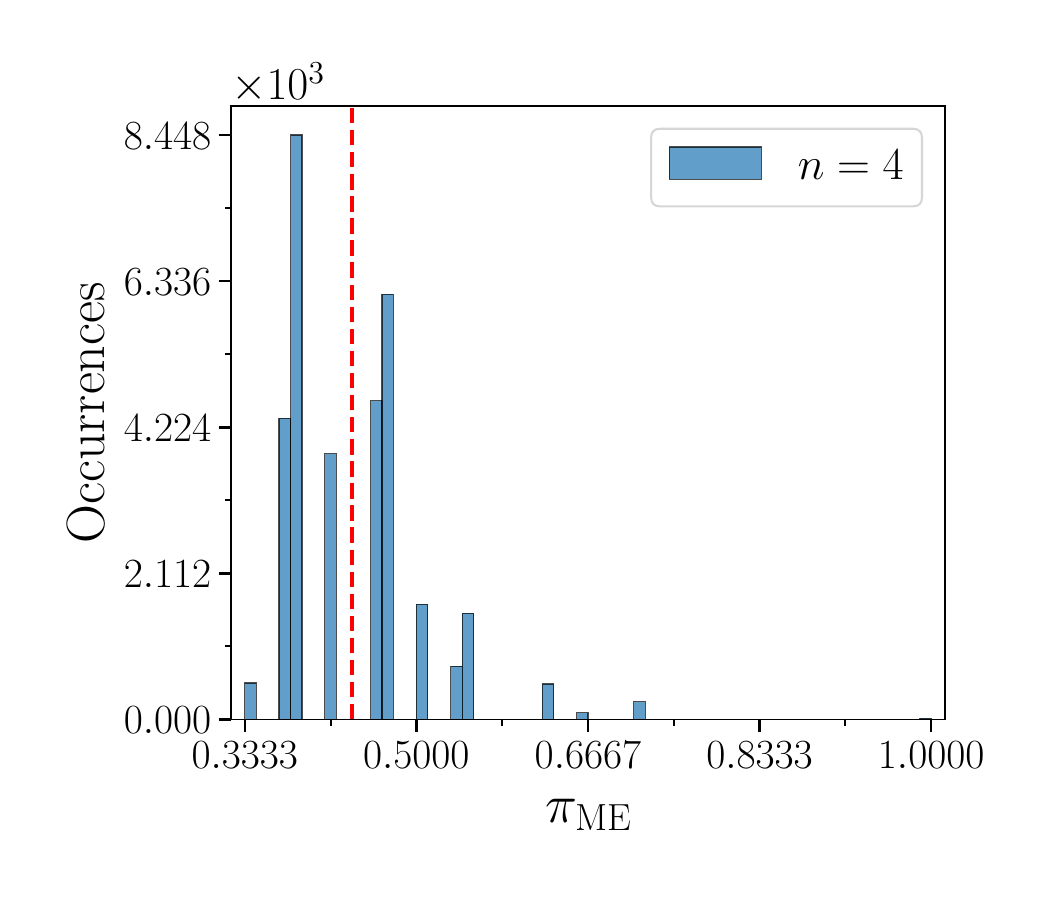}%
    }\hspace{0.03\textwidth}
    \subfloat[\label{fig:hist5}][]{%
        \includegraphics[width=0.3\textwidth]{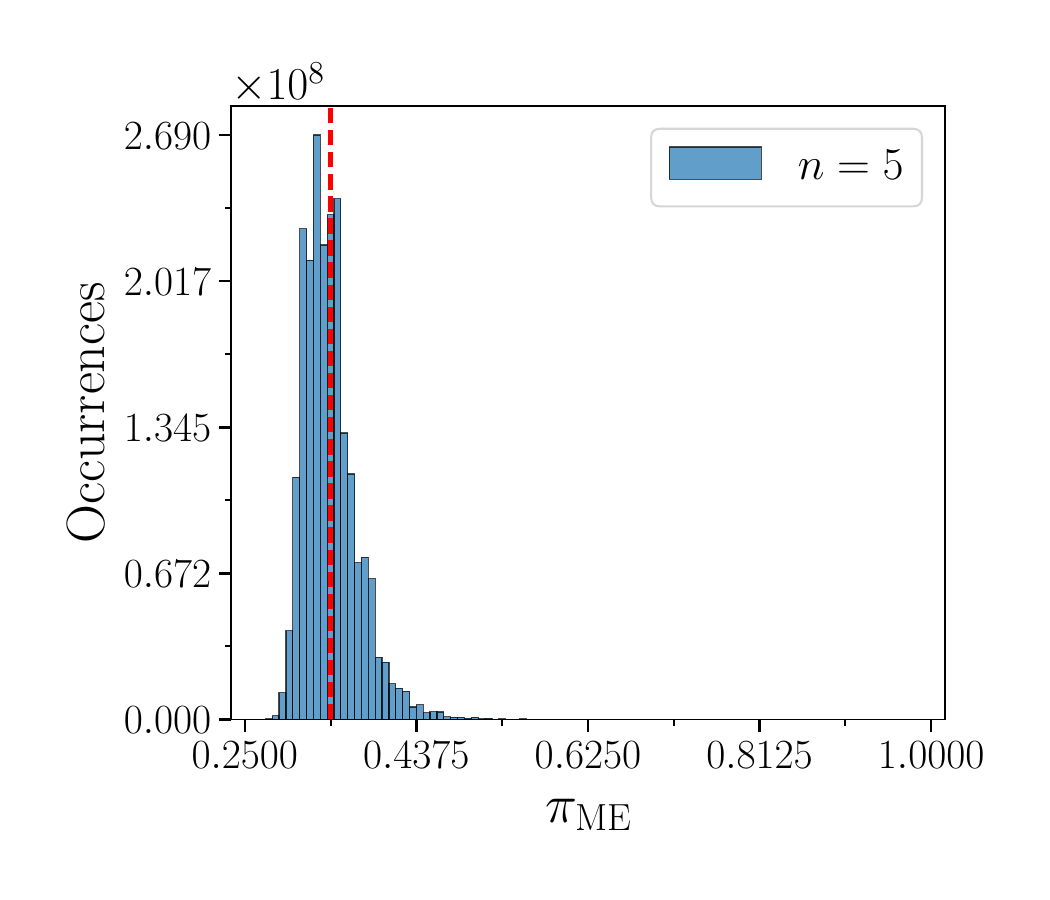}%
    }\\[1.5ex]

    \subfloat[\label{fig:U3}][]{%
        \includegraphics[width=0.3\textwidth]{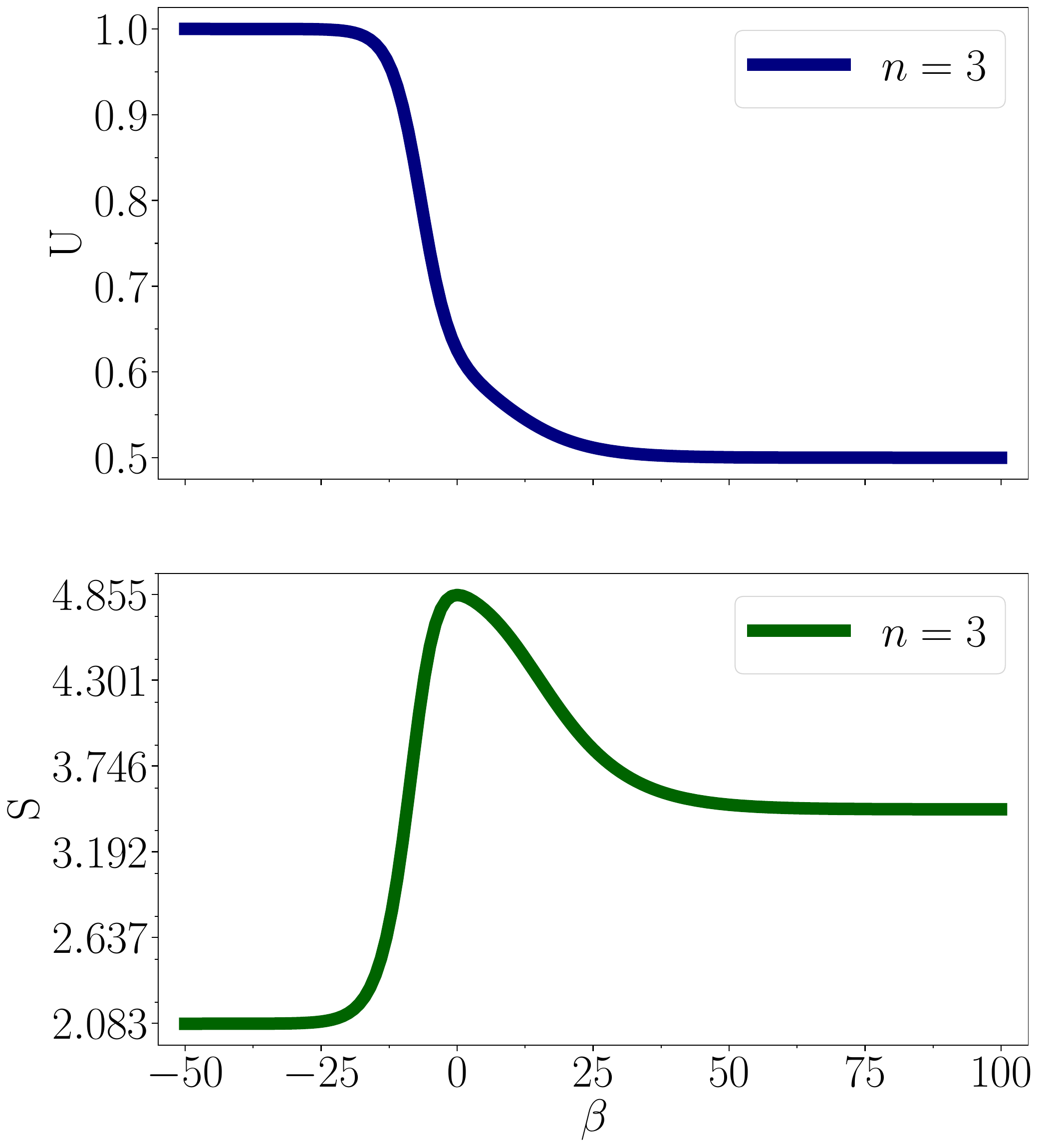}%
    }\hspace{0.03\textwidth}
    \subfloat[\label{fig:U4}][]{%
        \includegraphics[width=0.305\textwidth]{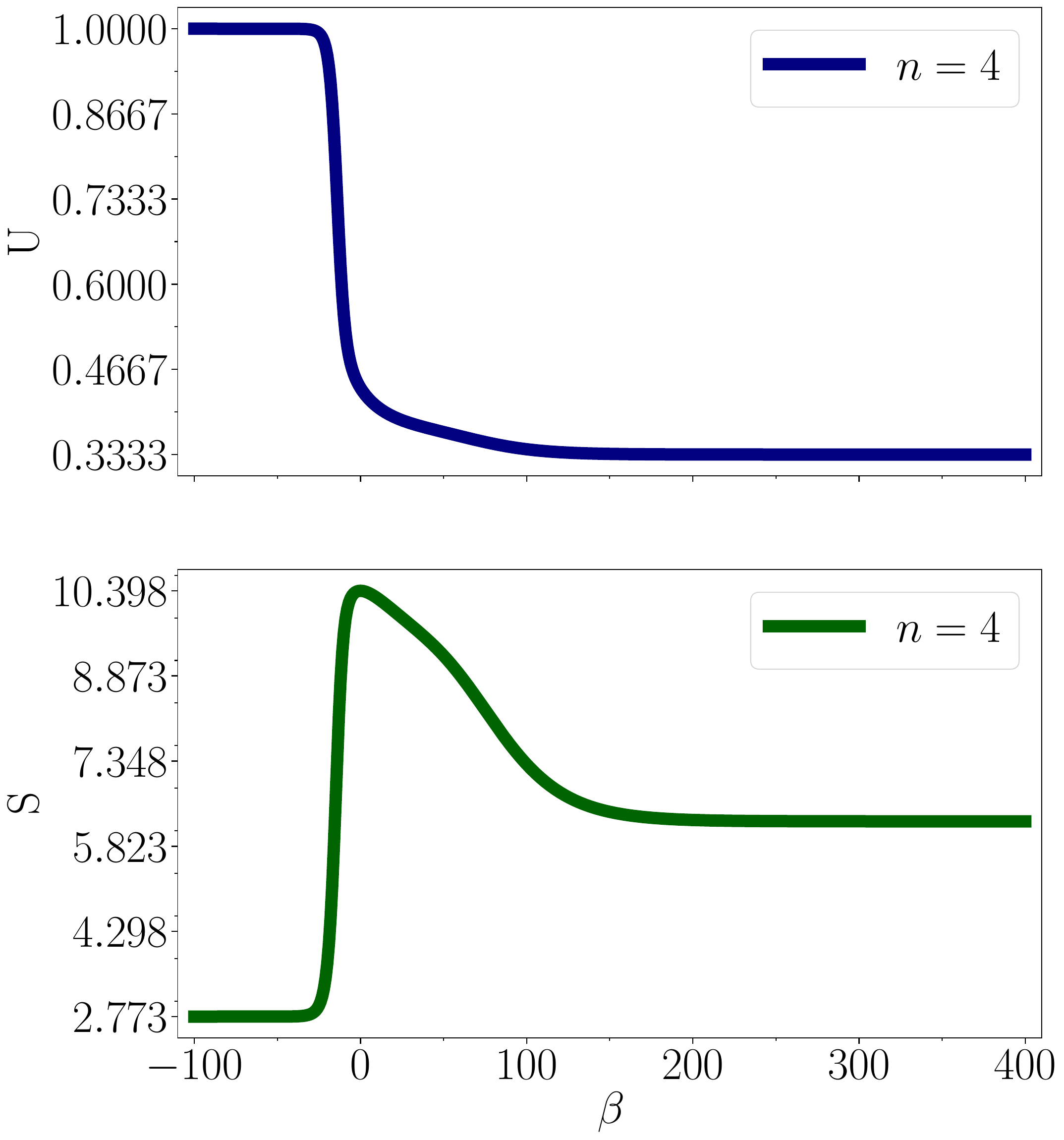}%
    }\hspace{0.03\textwidth}
    \subfloat[\label{fig:U5}][]{%
        \includegraphics[width=0.3\textwidth]{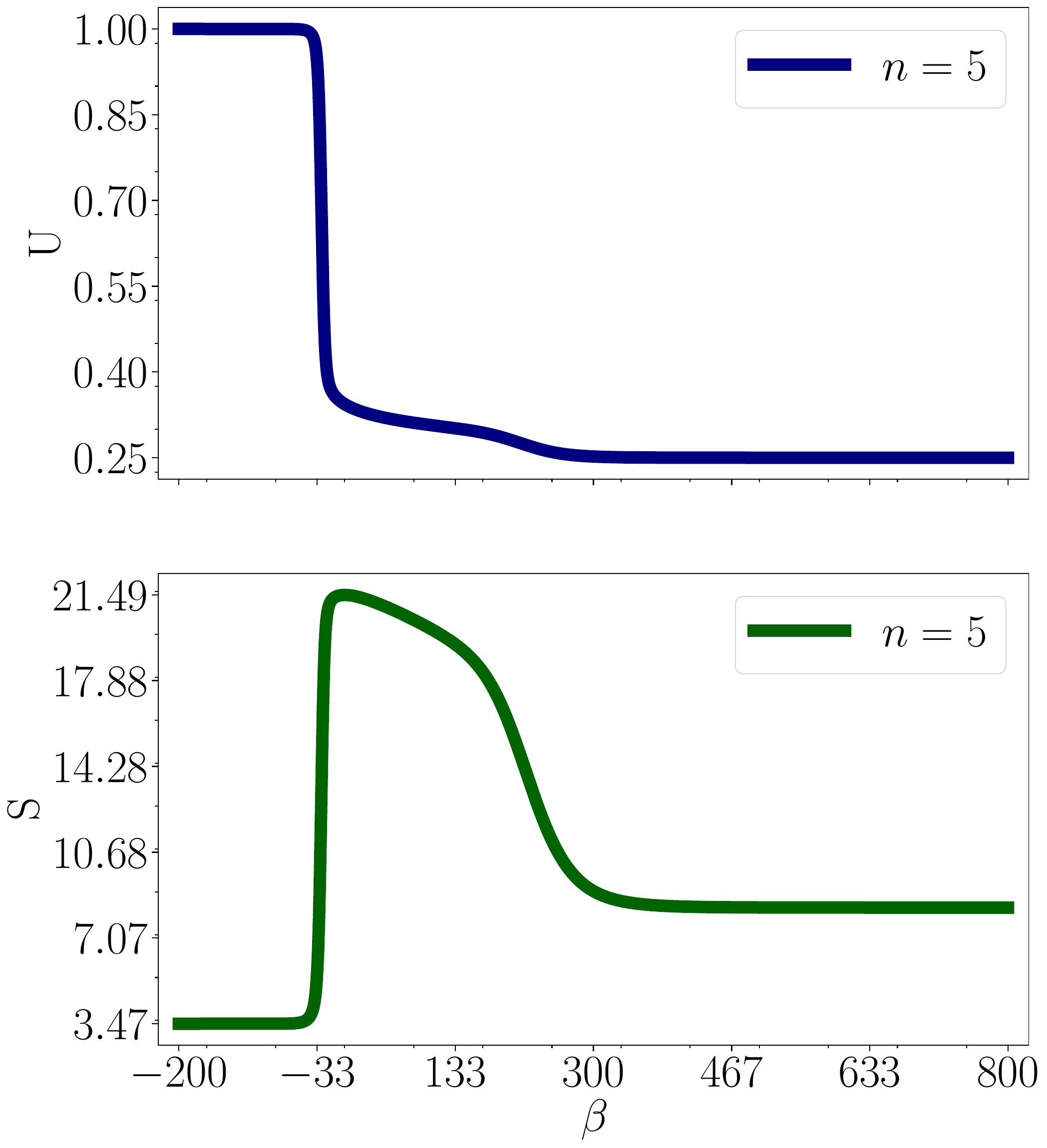}%
    }\\[1.5ex]

    \caption{\justifying Exact results for the hypergraph state sets of 3, 4, and 5 qubits.
    Top: distribution of multipartite entanglement $\pi_{\textrm{ME}}$, equivalent to the distribution of all microstates at $\beta = 0$: 
    (a) 3-qubit distribution, 
    (b) 4-qubit distribution, 
    (c) 5-qubit distribution. 
    Bottom: internal energy and entropy as a function of inverse temperature $\beta$: 
    (d) 3 qubits, $\beta \in [-50, 100]$; 
    (e) 4 qubits, $\beta \in [-100, 400]$; 
    (f) 5 qubits, $\beta \in [-200, 800]$. 
    }
    \label{fig:analytic-panel-345}
\end{figure*}

As mentioned in Sec.~\ref{sec:sec1}, the hypergraph state space is finite and contains $2^{2^n - 1}$ distinct elements for an $n$-qubit system. For $n=3,4,5$, the total number of states (respectively $128,\;32\,768,\; 2\,147\,483\,648$) remains computationally tractable for a complete enumeration. This enables the exact construction of the energy landscape and the associated thermodynamic observables, providing benchmark results for validating both the method and the simulated annealing approach later employed for larger systems.

For each system size, we systematically generated all possible sign configurations $\bm{s} = (s_0, s_1, \dots, s_{2^{n}-1})$, with $s_0=1$, and computed the corresponding multipartite entanglement potential $\pi_{\mathrm{ME}}(\bm{s})$ using Eq.~(\ref{eq:potential-unif}). The computations were performed on the High-Performance Computing cluster ReCaS~\cite{RECAS}. For $n=5$, the computation required several days using $50$ cores, with a data volume exceeding $650$~GB. From the complete enumeration, we constructed histograms of $\pi_{\mathrm{ME}}$ over the entire set of hypergraph states, shown in Fig.~\ref{fig:analytic-panel-345} (first row).

For $n=3$, the distribution of $\pi_{\mathrm{ME}}$ exhibits a peak at $\pi_{\mathrm{ME}}=1$, corresponding to separable states, a broad population at intermediate values, and a sharp peak at the minimum value $\pi_{\mathrm{ME}}=1/2$, corresponding to MMES. As the system size increases to $n=4$ and $n=5$, the distributions become increasingly concentrated and approach a quasi-continuous profile due to the exponential growth of configurations. The minimum values extracted from the distributions are $\pi^{(\min)}_{\mathrm{ME}}=1/2$ for $n=3$, $\pi^{(\min)}_{\mathrm{ME}}=1/3$ for $n=4$, and $\pi^{(\min)}_{\mathrm{ME}}=1/4$ for $n=5$. Thus, the cases $n=3$ and $n=5$ saturate the lower bound $\pi_{\mathrm{ME}}=1/2^{\lfloor n/2 \rfloor}$, confirming the existence of \emph{absolute maximal entangled} (AME) states, or
\emph{perfect} MMES, within the hypergraph state set. Conversely, the $n=4$ case does not saturate this bound due to frustration \cite{Higuchi_2000}, consistently with the known absence of AME$(4,2)$ states in four-qubit systems~\cite{Facchi_Lincei_2009,Facchi_frustartion_2010, Zyczkowsi_2015, Zyczkowsi_2018}. 

From the exact histograms of $\pi_{\mathrm{ME}}$, we obtain both the set of distinct values $\{\pi_\ell\}$ and their corresponding occurrences $\{f_\ell\}$. While this information fully characterizes the multipartite entanglement landscape and the MMES configurations for small system sizes, a complete enumeration rapidly becomes infeasible as $n$ increases. 

To overcome this limitation, we recast the problem into an equivalent classical statistical-mechanics framework, as explained in Sec.~\ref{sec:sec1}. Each hypergraph quantum state is associated with a configuration of classical binary spin system, the values $\pi_\ell$ play the role of energies $E_\ell$, and the occurrences $f_\ell$ define the density of states in~\eqref{eq:statedens} as
\begin{equation}
    P(E) = \sum_{\ell} f_{\ell}\, \delta(E - \pi_{\ell}).
\end{equation}
The histograms obtained from complete enumeration therefore correspond to the distribution of microstates at $\beta=0$, i.e., in the infinite-temperature limit ($T=\infty$), where all configurations are equally probable.

Within this framework, we construct the partition function~(\ref{eq:partition-function}) at inverse temperature $\beta$, which reads
\begin{equation}
    \mathcal{Z}(\beta) = \sum_{\ell} f_{\ell}\, e^{-\beta \pi_{\ell}}.
\end{equation}
This allows us to compute the thermodynamic observables, in particular the internal energy $U(\beta)$ by Eq.~\eqref{eq:internal-energy} and the entropy $S(\beta)$ by Eq.~\eqref{eq:entropy2}, as a function of the inverse temperature $\beta$. The resulting thermodynamic functions are shown in Fig.~\ref{fig:analytic-panel-345} (second and third rows).

Following the standard approach adopted for this class of models~\cite{Facchi_Statist_mechanics_2009,Facchi_Classical-statistical-mechanics_2010}, varying $\beta$ enables one to map the expectation value of the multipartite entanglement potential onto the purity axis. In this picture, the internal energy $U$ corresponds to the expectation value of $\pi_{\mathrm{ME}}$ [see Eq.~(\ref{eq:internal-energy})]. In the limit of large positive $\beta$, $U$ approaches the lower bound~$\pi^{(\min)}_{\mathrm{ME}}$ of the potential, thus identifying the MMES configurations.

The extension to negative $\beta$ is also physically meaningful. While the sign of $\beta$ can be regarded as a convention in the analogy with classical spin systems (e.g., ferromagnetic versus antiferromagnetic interactions), it provides access to complementary regimes of the optimization landscape. For $\beta>0$, the internal energy decreases monotonically, corresponding to the minimization of the cost function. Conversely, for $\beta<0$, the same formalism leads to the maximization of $U$, driving the system toward configurations with maximal $\pi_{\mathrm{ME}}$, i.e., separable states, at $\pi^{(\max)}_{\mathrm{ME}}=1$. This regime is particularly useful as a consistency check, since its properties are well understood.

The numerical results for small system sizes confirm this behavior. The internal energy $U(\beta)$ decreases monotonically with increasing $\beta$ and, in the limit $\beta \to \infty$ ($T \to 0^+$), the system converges to MMES configurations. In the negative-$\beta$ region, the system instead approaches separable states. 

The entropy $S(\beta)$ quantifies the number of accessible configurations at each temperature. At $\beta=0$, the entropy attains its maximum value, equal to the logarithm of the total number of hypergraph states, in agreement with Table~\ref{tab:sub-sets}. For $\beta>0$, the entropy decreases as the system explores low-energy (highly entangled) configurations and eventually reaches a plateau in the zero-temperature limit. This plateau encodes the logarithm of the number of MMES, yielding $n_{\mathrm{MMES}}=32$ for $n=3$, $528$ for $n=4$, and $4224$ for $n=5$. An analogous behavior is observed for negative $\beta$, where the entropy plateau corresponds to the number of separable states. 
Overall, $S(\beta)$ captures the continuous interpolation between the two extremal regimes, highlighting how temperature drives the system across different entanglement phases, from a disordered (random) phase to the two ordered phases associated with maximally entangled and fully separable configurations.

\section{Thermodynamics for larger systems: Annealing Methods}\label{sec:sec3}

The exact enumeration approach employed in Sec.~\ref{sec:sec2}, while providing complete characterization for $n \leq 5$, becomes computationally infeasible for $n \geq 6$ due to the (doubly) exponential growth of the set cardinality~\eqref{eq:N_hyper} (see Table \ref{tab:sub-sets}). 
To address this computational challenge, we employ a simulation procedure based on a {Parallel Tempering} (PT) algorithm \cite{Tempering_1986, Geyer_1991, Marinari/Parisi_1992, Tempering_1996, Falcioni_1999, Grigera_2001, Earl_2005}that efficiently samples the configuration space (for technical details and comments, see Appendix~\ref{sec:appA}). This approach allows for an approximate evaluation of the internal energy associated with the hypergraph states without requiring a full traversal of the state space. Substantial computational resources and advanced sampling strategies are essential to reduce statistical errors and improve convergence, thereby enabling the reliable generation of thermodynamic observables. Several key parameters must be carefully specified to ensure the effectiveness of the simulation: the range of inverse temperatures $\beta$, the number of annealing steps per $\beta$, the number of independent runs of the full protocol and the number of replicas of system, which has to be the same number of $\beta$'s values, see Appendix~\ref{sec:appA}.  

The $\beta$ range must be broad enough for the system to approach its asymptotic thermodynamic behavior in the limits $\beta \rightarrow \pm\infty$. Since the exact onset of these regimes is not known a priori, preliminary simulations varying the bounds of the range of $\beta$ can help identify thresholds where saturation occurs. The number of annealing steps should be sufficient to allow the replicas of system at any temperatures to sample a representative subset of the accessible configuration space (i.e., the $2^{2^n - 1}$ spin configurations defining the set), while remaining computationally tractable. Similarly, the number of independent repetitions must balance statistical significance with computational cost, as it directly affects the time of runs and all the parallelization protocols.

The simulation parameters used for each system size are summarized in Table \ref{tab:parameter-therm-simulation}.
\begin{table*}[t]
\begin{ruledtabular}
\begin{tabular}{ccccc}
 \textbf{Qubits ({n})}  & \textbf{Replicas/number of betas} & \textbf{$\beta$ range} & \textbf{Annealing Steps} & \textbf{Independent runs}  \\ 
         \hline
         3  & $54  = (27+27)$ & $[-50, 100]$ &  $100{,}000$ & $50$  \\ 
         4 & $74 = (45 +29)$  & $[-100, 400]$  & $250{,}000$  & $50$ \\
         5 & $108 = (66 + 42)$  & $[-200, 800]$  & $1{,}000{,}000$  & $50$  \\
         6 & $220 =(89 + 131)$  & $[-200, 1000]$  & $ 2{,}000{,}000$ & $50$  \\
         7 & $393 = (204 + 189)$ & $[-500, 15000]$  & $5{,}000{,}000$ & $100$ \\
\end{tabular}
\end{ruledtabular}
\caption{\justifying Key control parameters for the simulations of the internal energy using parallel tempering. The second column reports the number of replicas, i.e., the number of inverse temperatures ($\beta$ values), expressed as the sum of two contributions corresponding to the replicas used in the positive- and negative-$\beta$ regimes. See text.}
\label{tab:parameter-therm-simulation}
\end{table*}
The simulations are performed as follows. For each independent run, we apply the annealing procedure starting from $\beta = 0$, treating the $\beta < 0$ and $\beta > 0$ regions separately. Any replica in the simulation runs at a different value of $\beta$ in the range chosen and the associated internal energy is estimated by averaging the values of the potential of multipartite entanglement in Eq.~(\ref{eq: multpotentialdef}) over the last $\sim25\%$ of the annealing steps, thereby suppressing transient effects. This protocol produces, for each independent run, a dataset $U(\beta)$ over the chosen discrete set of $\beta$ values, i.e.\ a numerical representation of thermodynamic function $U(\beta)$ for that run.
The entropy is extracted from the internal energy according to Eq.~(\ref{eq:entropy2}). To avoid error propagation and possible correlations between $U(\beta)$ and $S(\beta)$, we apply Eq.~(\ref{eq:entropy2}) independently to each $U(\beta)$ dataset as a function of $\beta$. In this way, from every independent realization of $U(\beta)$ we obtain a corresponding dataset $S(\beta)$. Repeating this procedure for all independent runs (50 in our analysis), we obtain ensembles of datasets for both $U(\beta)$ and $S(\beta)$ over the suitable  discrete interval of $\beta$ chosen.
Finally, for each selected value of $\beta$, we collect the corresponding values of internal energy and entropy from all datasets, compute their averages, and estimate the statistical error. At the end of this protocol, we obtain two distinct plots: one for the internal energy $U(\beta)$ and one for the entropy $S(\beta)$, both as functions of $\beta$, with associated error bars at each point.

\section{Validation of the method and MMES counting}\label{sec:sec4}

In order to validate our results, we hinge upon distinctive properties of the thermodynamic functions: both the internal energy and entropy curves possess particular points whose analytical values are exactly known. These reference points serve as natural benchmarks, allowing us to directly quantify the accuracy of the annealing-based estimates. They are defined as follows: 

\begin{enumerate}
\item The energy at $\beta = 0$, denoted as $U(0)$, corresponds to the expectation value of the distribution of the potential of multipartite entanglement over the complete ensemble of pure $n$-qubit hypergraph states \cite{Facchi_Lincei_2009, Trotta_2026}. This quantity can be directly evaluated from analytical calculations for each system size, as detailed in Appendix~\ref{sec:appC}.
\item The energy in the limit $\beta \rightarrow +\infty$, denoted by $U(+\infty)$, corresponds to the lower bound of the potential of multipartite entanglement. This quantity is known exactly up to $n = 6$ qubits.

\item The entropy in the limit $\beta \to -\infty$, corresponding to the logarithm of the number of separable states within the ensemble of pure hypergraph states. For a system of $n$ qubits to be separable, under the constraints imposed by the hypergraph state  set, each qubit can only occupy one of the two configurations $|0\rangle \pm |1\rangle$, up to a normalization factor. Consequently, the total number of possible separable configurations is exactly $n_{\textrm{SEP}} = 2^n$. In the following, we directly compare the number of separable states with the bounds associated to the estimated number of states, as inferred from the exponential of the entropy in the asymptotic regime $\beta \to -\infty$, considering the statistical errors.

\item The entropy in the limit $\beta \to +\infty$ corresponds to the logarithm of the number of maximally entangled states within the ensemble of pure hypergraph states. For $n=3,4,$ and $5$ qubits, we enumerated the entire set and obtained the exact numbers of MMES, namely $32$, $528$, and $4224$, respectively. For $n=6$ qubits, it is known that maximum entanglement is attained by stabilizer states~\cite{Scott_2004}. Within the set of hypergraph states, {all stabilizer states are graph states (up to local Pauli equivalence) ~\cite{Macchiavello_2026}.} 
This implies that, for $n=6$, the search for MMES can be restricted to the set of graph states rather than the full set of hypergraph-state. Consequently, an exhaustive enumeration remains feasible, since the total number of graph states, {comprehensive of local Pauli transformations}, is only of the order of a few million. From the average purity calculation, we find for $n = 6$ a total of $8448$ maximally multipartite entangled states.
\end{enumerate}

The results obtained from the numerical simulations have been tabulated with the corresponding analytical values, when they are known. More precisely, the comparisons for separable states, MMES, and internal energies are shown in 
Table \ref{tab:sep-values}, Table \ref{tab:mmes-values} and Table \ref{tab:mean_value&lowest_value U}, respectively. The agreement is excellent. 
 
\begin{table*}[t]
\centering
\begin{ruledtabular}
\begin{tabular}{ccccc}
 \textbf{Qubits ($n$)} & \textbf{$S_{\mathrm{SEP}}^{th}$} & \textbf{$S_{\mathrm{SEP}}^{num}$} & \textbf{$n_{\mathrm{SEP}}^{th}$} & \textbf{ Bounds $n_{\mathrm{SEP}}^{num}$} \\
 \hline
 3 & 2.07944 & $2.083 \pm 0.003$ &  8   & $8.00 \leq n_{\mathrm{SEP}} \leq 8.05$ \\
 4 & 2.77259 & $ 2.773 \pm  0.004$ & 16  &  $ 15.95\leq n_{\mathrm{SEP}} \leq16.07 $\\
 5 & 3.46574 & $3.467 \pm0.005$ & 32  & $ 31.89 \leq n_{\mathrm{SEP}} \leq 32.21 $ \\
 6 & 4.15888 & $4.171 \pm 0.012 $ & 64  & $  64.0\leq n_{\mathrm{SEP}} \leq65.6$\\
 7 & 4.85203 & $ 4.859  \pm 0.015 $ & 128 & $ 127\leq n_{\mathrm{SEP}} \leq 131 $\\
\end{tabular}
\end{ruledtabular}
\caption{\justifying Comparison between analytical and numerical results for separable states (SEP).
Second column: $S_{\mathrm{SEP}}^{th} = \log(n_{\mathrm{SEP}}^{th})$ is the analytical value of the entropy at temperature $T = 0^-$. 
Third column: $S_{\mathrm{SEP}}^{num}$ is the numerical estimate, with statistical error $= 1\sigma$, of the entropy at temperature $T = 0^-$.
Fourth column: $n_{\mathrm{SEP}}^{th} = 2^{n}$ is the analytical value of the number of separable states.
Fifth column: numerical estimate, with statistical error, of the number of separable states.}
\label{tab:sep-values}
\end{table*}

\begin{table*}[t]
\centering
\begin{ruledtabular}
\begin{tabular}{ccccc}
 \textbf{Qubits ($n$)} & \textbf{$S_{\mathrm{MMES}}^{th}$} & \textbf{$S_{\mathrm{MMES}}^{num}$} & \textbf{$n_{\mathrm{MMES}}^{th}$} & \textbf{Bounds $n_{\mathrm{MMES}}^{num}$} \\
 \hline
 3 & 3.46574 & $3.467 \pm 0.002$ & 32   & $31.97 \leq n_{\mathrm{MMES}} \leq 32.11$ \\
 4 & 6.26910 & $6.272 \pm  0.003$ & 528   & $527.7 \leq n_{\mathrm{MMES}} \leq 531.2$ \\
 5 & 8.34854 & $8.351 \pm 0.004$ & 4224   & $4,220 \leq n_{\mathrm{MMES}} \leq 4,251$ \\
 6 & $\ge 9.04169$ & $9.037 \pm 0.006$ &  $\ge 8448$   & $ 8,358  \leq n_{\mathrm{MMES}} \leq 8,466 $\\
 7 & - & $ 14.79 \pm 0.09 $ & -   & $ 2,434,120  \leq n_{\mathrm{MMES}} \leq 2,898,520 $ \\
\end{tabular}
\end{ruledtabular}
\caption{\justifying Comparison between analytical and numerical results for maximal multipartite entangled states (MMES).
Second column: $S_{\mathrm{MMES}}^{th} = \log(n_{\mathrm{MMES}}^{th})$ is the analytical value of the entropy at temperature $T = 0^+$. 
Third column: $S_{\mathrm{MMES}}^{num}$ is the numerical estimate, with statistical error $= 1\sigma$, of the entropy at temperature $T = 0^+$.
Fourth column: $n_{\mathrm{MMES}}^{th}$ is the analytical value of the number of MMES (see text).
Fifth column: numerical estimate, with statistical error, of the number of MMES.
All the analytical results for $n = 6$ qubits are restricted to the graph subset.}
\label{tab:mmes-values}
\end{table*}

\begin{table*}[t]
\centering
\begin{ruledtabular}
\begin{tabular}{ccccc}
 \textbf{Qubits ($n$)} & \textbf{$U_{th}(0)$} & \textbf{$U_{num}(0)$} & \textbf{$U(+\infty)$}\\
 \hline
 3 & $5/8 = 0.625 $  & $0.6251 \pm 0.0005$ &  $1/2$ \\
 4 & $7/16 = 0.4375$ & $0.4374 \pm 0.0003$ & $1/3$\\
 5 & $11/32 = 0.34375 $  & $0.34379\pm 0.00007$ & $1/4$ \\
 6 & $ 15/64 = 0.234375 $ & $ 0.234363 \pm 0.000015$ & $1/8$ \\
 7 & $ 23/128 = 0.1796875$   & $0.179680\pm 0.000008$ &  $0.13504766\pm 2.6\times10^{-7}$ \\
\end{tabular}
\end{ruledtabular}
\caption{\justifying
Comparison between analytical and numerical estimates of the internal energy.
Second column: analytical value of the internal energy at $\beta = 0$ ($T = \infty$) (see Appendix \ref{sec:appC}).
Third column:  numerical estimate, with statistical error $= 1\sigma$, of the internal energy at $\beta = 0$ ($T = \infty$).
Fourth column: internal energy at the maximum of $\beta$ achieved (see Table \ref{tab:parameter-therm-simulation}); for $n = 7$, see Eq.\ (\ref{eq:pi_7}).}
\label{tab:mean_value&lowest_value U}
\end{table*}

\subsection{$n = 3$, $4$, and $5$ qubits}\label{sec:sec345}
\begin{figure*}[t]
    \centering

    \subfloat[]{%
        \includegraphics[width=0.3\textwidth]{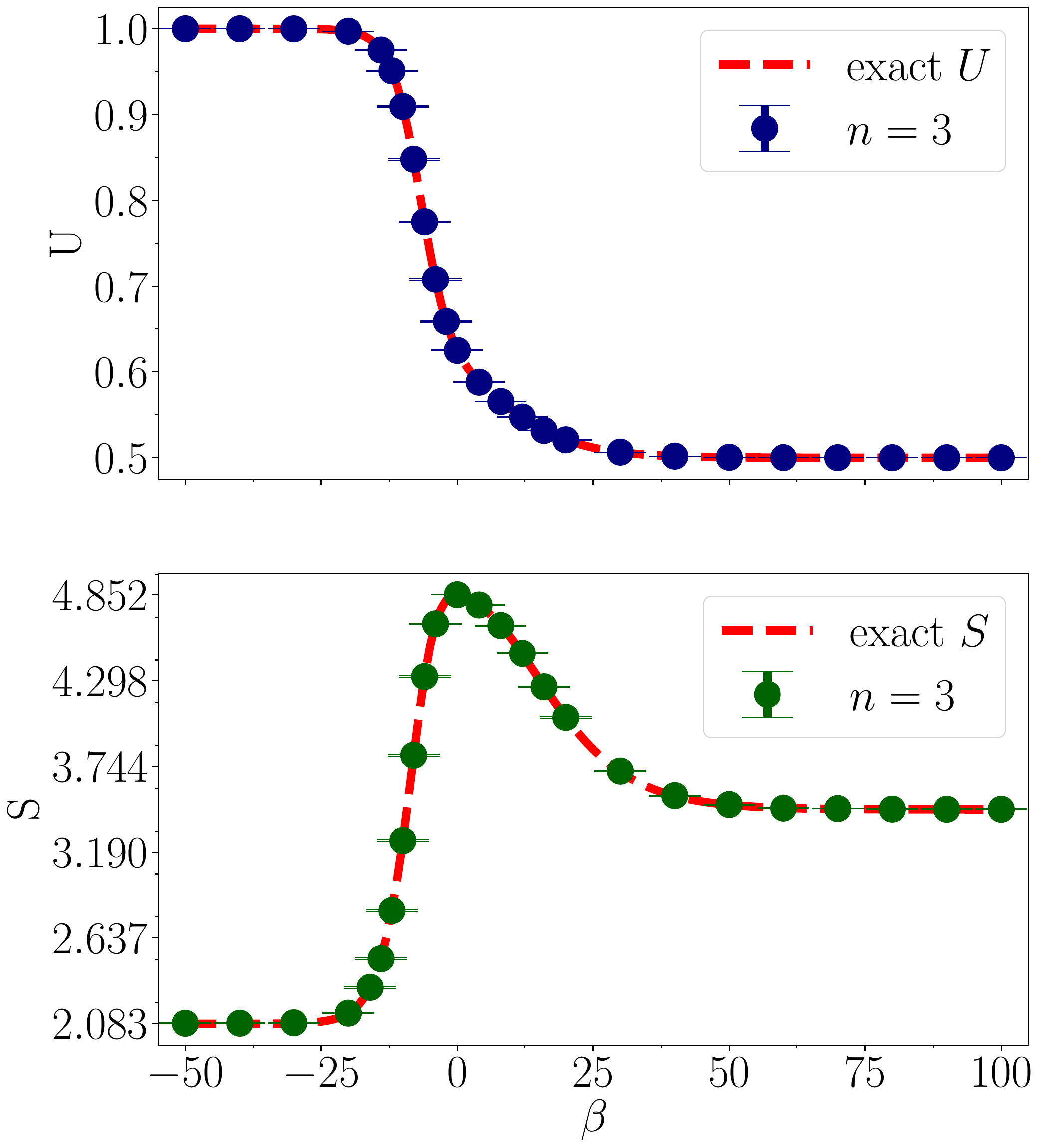}%
    }\hspace{0.01\textwidth}
    \subfloat[]{%
        \includegraphics[width=0.31\textwidth]{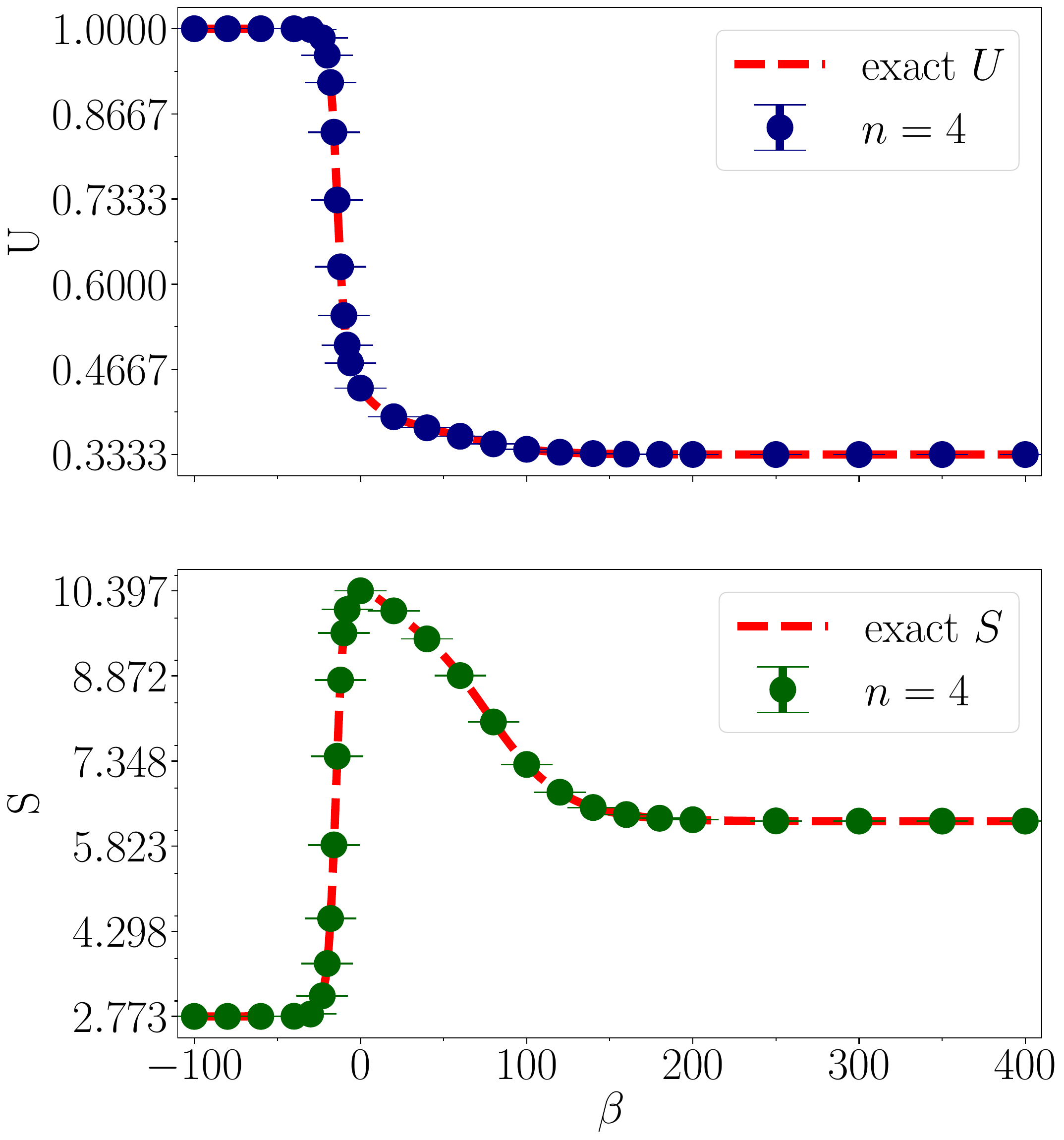}%
    }\hspace{0.01\textwidth}
    \subfloat[]{%
        \includegraphics[width=0.3\textwidth]{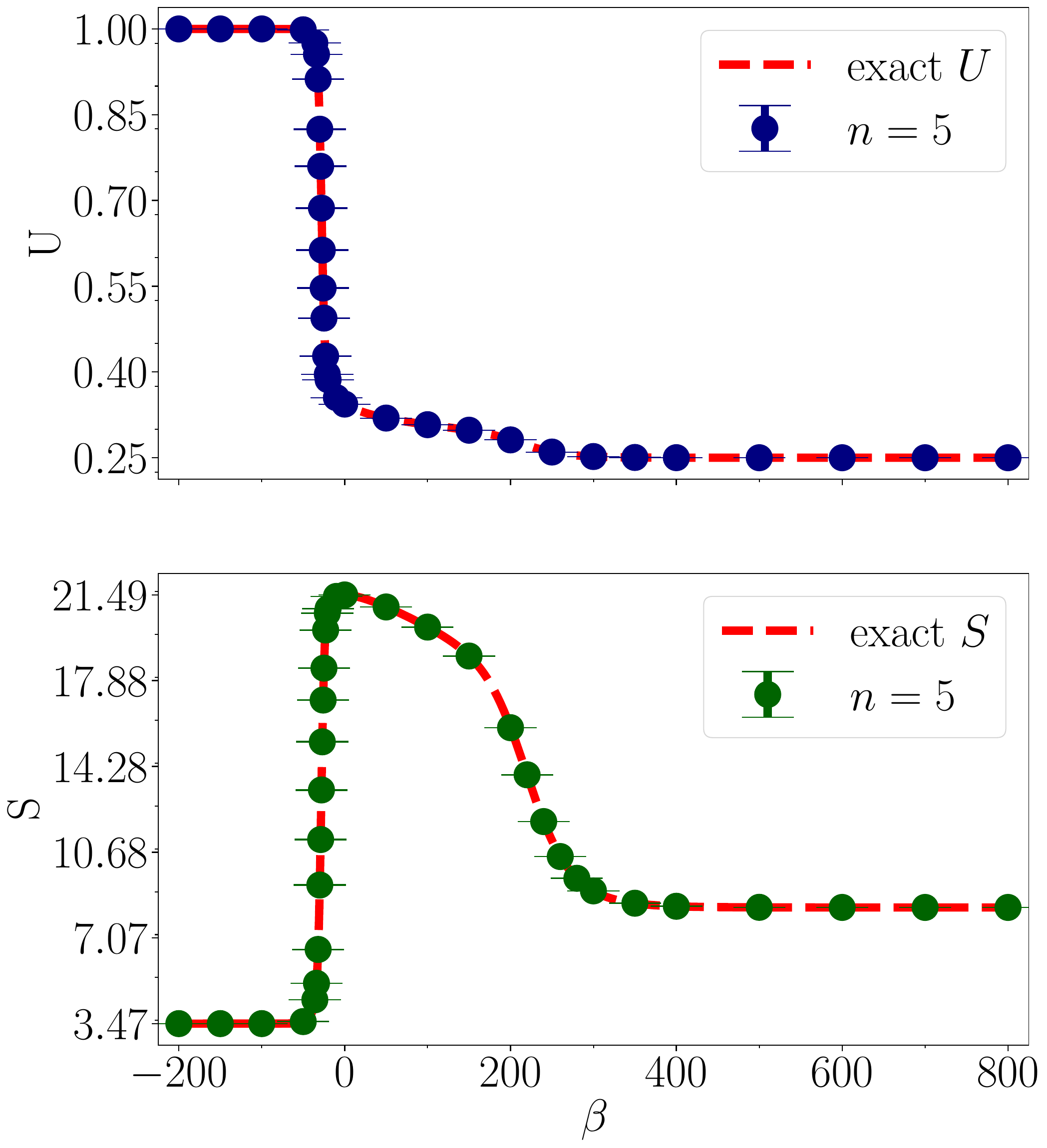}%
    }\hspace{0.01\textwidth}
    \\[1.5ex]
    
    \caption{\justifying Comparison between the simulated internal energy and entropy (points with error bars) and those exactly computed in Sec.\ \ref{sec:sec2}. (a) 3-qubit, (b) 4-qubit, (c) 5-qubit. The dashed curves are those shown in Fig.\ \ref{fig:analytic-panel-345}.}
    \label{fig:comp_345}
\end{figure*}

For $n = 3$, $4$, and $5$ qubits, the numerical results for the thermodynamic functions have been compared with the exact predictions obtained from the complete enumeration of the corresponding sets, outlined in Sec.\ \ref{sec:sec2}.
A visual comparison of the thermodynamic functions shows close agreement between the numerical results and the exact values, and is displayed in Fig.\ \ref{fig:comp_345}, where the dashed curves are those shown in Fig.\ \ref{fig:analytic-panel-345}, obtained from the explicit computation of the full sets of hypergraph states in Sec.\ \ref{sec:sec2}.

Figure \ref{fig:comp_345} demonstrates that, across all these system sizes, the simulated data closely match the exact calculations. The close correspondence underscores the accuracy and robustness of the annealing protocol in reproducing the key thermodynamic features, even as the dimensionality of the system increases.

It is worth noting that for $n=4$ the minimum 1/4 cannot be attained, as the MMES has minimum average purity 1/3 (internal energy $U(+\infty)$, see Table \ref{tab:mean_value&lowest_value U}) \cite{Higuchi_2000, Facchi_Lincei_2009}.
This is the first occurrence of \emph{quantum entanglement frustration} \cite{Facchi_frustartion_2010}, and is due to the fact that the requirement that purity be minimal cannot be fulfilled for all bipartitions in Eq.\ (\ref{eq: multpotentialdef}). Interestingly, this feature is not manifest in Fig.\  \ref{fig:comp_345}. Since the cases $n=3,5$ and $6$ are not frustrated, the next instance of frustration will occur for $n=7$ and will be thoroughly discussed in Sec.\ \ref{sec:sec7only}. We will see that this feature will make the numerics and the analysis highly nontrivial.

\subsection{$n = 6$ qubits}\label{sec:sec6only}

It is clear from Table \ref{tab:sub-sets} that the total number of states of 6 qubits is too large to perform a complete numerical exploration. Nonetheless, as mentioned before, 
the absence of quantum frustration at 6 qubits \cite{Facchi_frustartion_2010} allows the existence of perfect MMES,(AME(6,2)), \cite{Facchi_Lincei_2009, Zyczkowsi_2015}.

Moreover, all maximal entangled states found with the annealing protocols are graph-states. We thus enumerated all graph-states and found a total of $8448$  AME(6,2), in agreement with the thermodynamics results, see \cite{Scott_2004} and Table \ref{tab:mmes-values}.

\begin{figure*}[t]
    \centering

     \subfloat[]{%
        \includegraphics[width=0.7\textwidth]{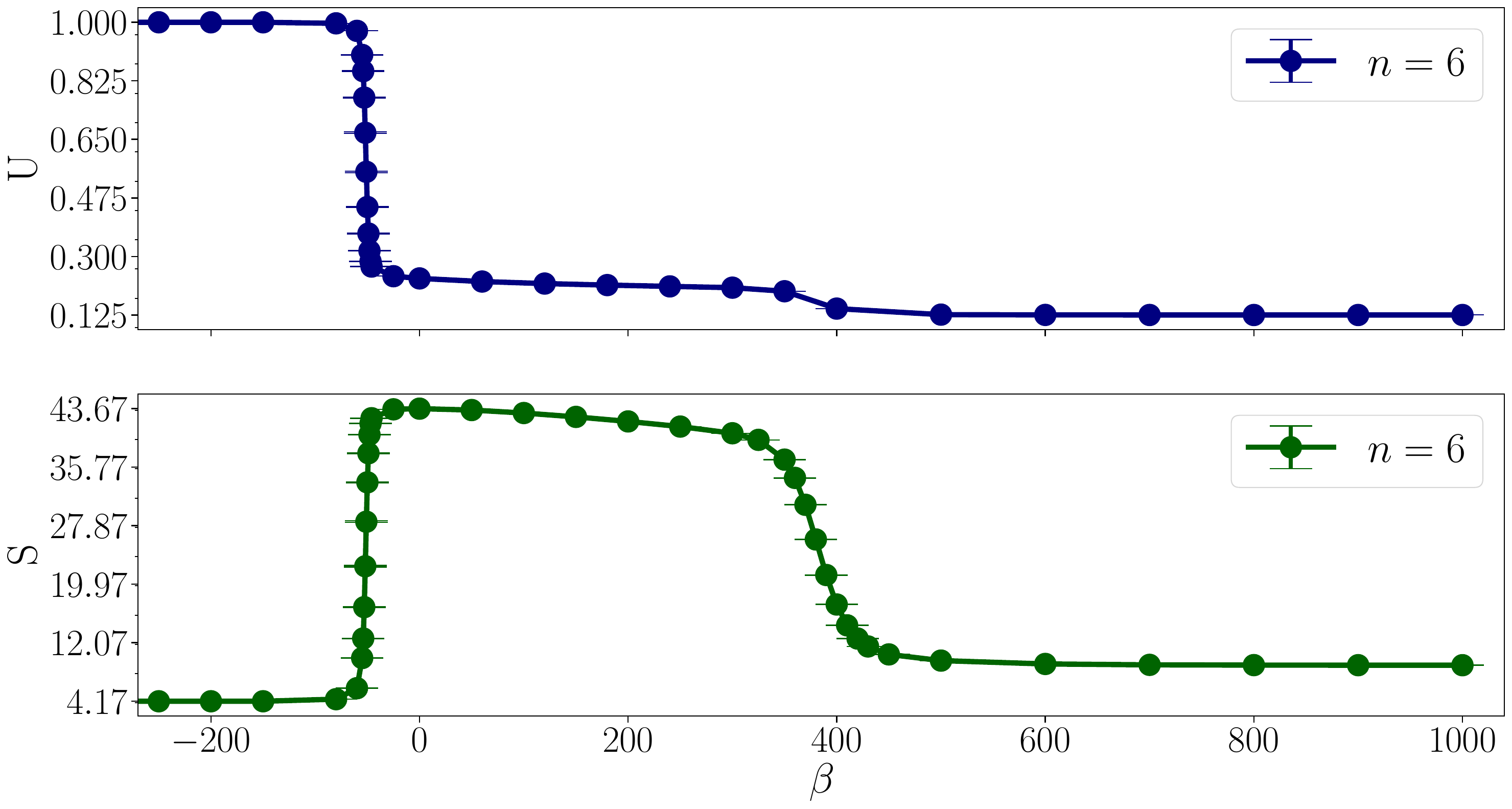}%
    }\hspace{0.01\textwidth}
       
    \caption{\justifying 
Internal energy (top) and entropy (bottom) for 6 qubits. }
    \label{fig:numerical-panel6}
\end{figure*}

The internal energies and entropies at $\beta = 0$ and $\beta = \pm \infty$ are given in Tables \ref{tab:sep-values}, \ref{tab:mmes-values} and \ref{tab:mean_value&lowest_value U} and are in excellent agreement with the numerical exploration. The case $n=6$ can be viewed as a complete, final validation of the numerical methods we used.
Internal energy and entropy are shown in Fig.\  \ref{fig:numerical-panel6}.

\subsection{$n = 7$ qubits}\label{sec:sec7only}

\begin{figure*}[t]
    \centering
    
    \subfloat[]{%
        \includegraphics[width=0.7\textwidth]{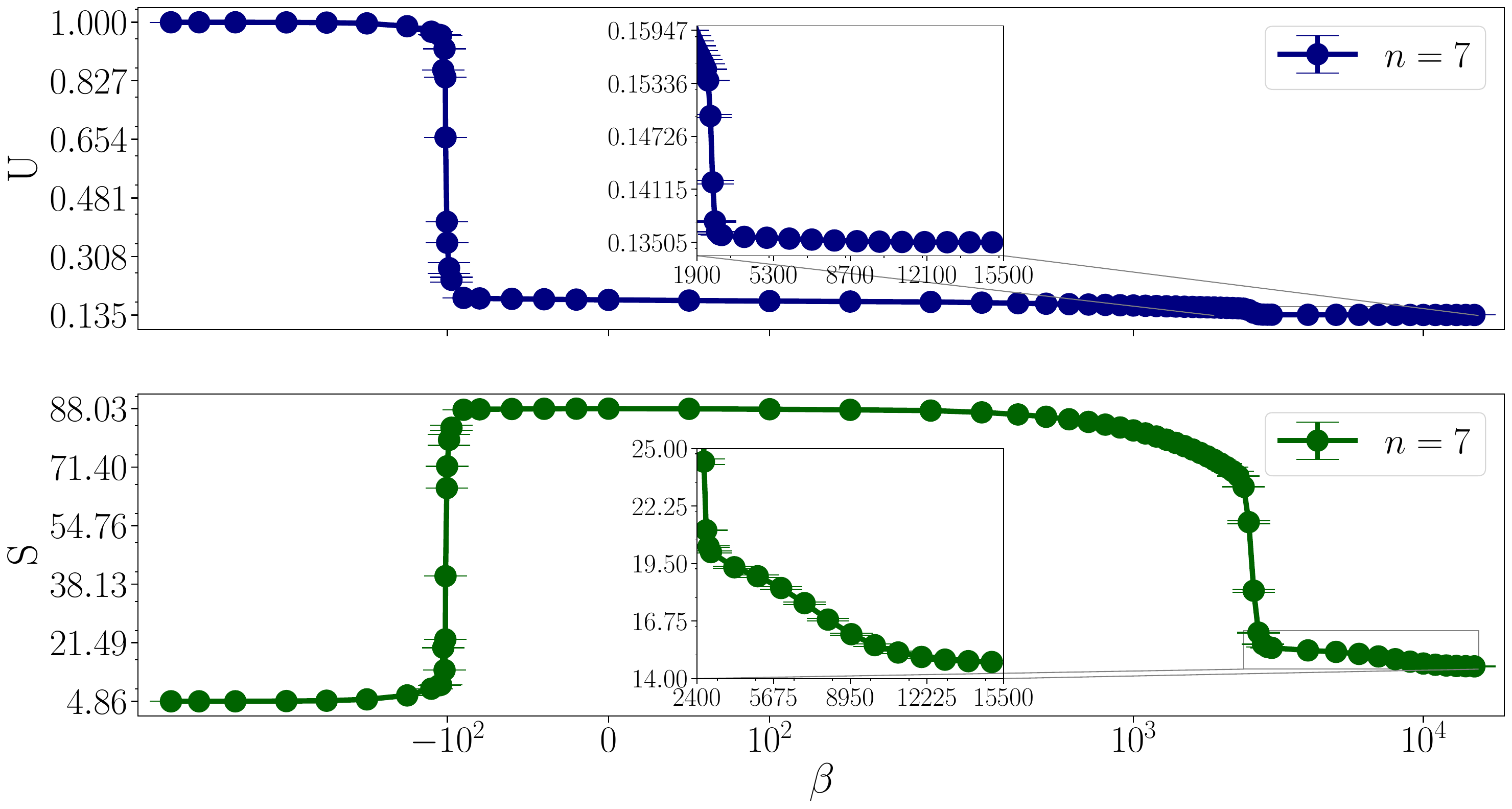}%
    }\hspace{0.01\textwidth}
   
    \caption{\justifying Internal energy (top) and entropy (bottom) for 7 qubits. The thermodynamic functions have been plotted using a log-scale on the $\beta-$axis for making evident the behavior in the negative $\beta$ region. 
     }
    \label{fig:numerical-panel7}
\end{figure*}

A significant challenge arises in the case of $n = 7$ qubits, which is expected to persist for larger system sizes, $n > 7$. The set of  hypergraph quantum states exhibits a pronounced energy barrier, separating the MMES, which we identify as ground-state candidates, from the typical values of the averaged purities, and hence of the internal energy. See Fig.\ \ref{fig:numerical-panel7}.
A detailed analysis reveals a generally low connectivity among the most entangled configurations. This feature can be characterized by exploiting their specific structure. Indeed, each  hypergraph state~(\ref{eq:unifstate}) is uniquely identified by the set ${\pm 1}$ associated with its relative phases. The component-wise difference between two such configurations can therefore be naturally mapped onto their Hamming distance. Using this metric, for the $n=7$ case, we find that the most entangled states in the set are highly separated, with an average Hamming distance $d \ge 60$, over a length $N = 2^{n} = 128$ of the sign vectors $\bm{s}$.

This finding clearly explains the difficulty in reaching MMES configurations through annealing. The simulated annealing procedure employed here is based on single-spin (single-phase) flip updates within the set. In the high-$\beta$ regime, however, the probability of escaping from local minima becomes negligibly small, making the exploration of distant regions of configuration space extremely unlikely.
To overcome this critical slowing down, for the $n = 7$ case only, we add to Parallel Tempering a preliminary annealing stage in which configurations that are not MMES are periodically randomized. This preparatory phase significantly enhances the probability of observing MMES configurations in the high-$\beta$ regime and accelerates convergence toward the most entangled states, partially breaking the barrier we observe for the internal energy plot.
The $n=7$ case is especially significant in view of the exponential growth of the set cardinality and the emergence of {entanglement frustration} \cite{Facchi_frustartion_2010,Scott_2004,Rains_1999}.

Due to frustration, the multipartite entanglement potential defined in Eq.~(\ref{eq: multpotentialdef}) does not attain the mathematical lower bound,
but has instead a higher, nontrivial minimum
\begin{equation}
\pi_{\mathrm{ME}}^{\min} > \frac{1}{2^{\lfloor n/2 \rfloor}},
    \label{eq:pi_ME}
\end{equation}
due to a ``competition" among different bipartitions. By employing the annealing-based approach described above, we find
\begin{equation}
    \pi_{\mathrm{ME}}^{\min} = 0.13504464 > \frac{1}{8}
    \label{eq:pi_7}
\end{equation}
and
\begin{equation}
    2{,}434{,}120 \leq n_{\mathrm{MMES}} \leq 2{,}898{,}520 .
    \label{n_mmes7}
\end{equation}

These findings provide a quantitative characterization of the first nontrivial frustrated regime. Although the $n=4$ case is already frustrated (MMES have internal energy 1/3 and not 1/4, see Sec.\ \ref{sec:sec2}), its small size does not allow the impact of quantum frustration to be fully appreciated. By contrast, the $n=7$ case represents the first instance in which frustration manifests itself in a quantitatively significant way. The annealing-based numerical approach enables us to identify for the first time, to the best of our knowledge, the physical minimum of the multipartite entanglement potential within the hypergraph state set. Furthermore, it provides an estimate of the size of the class of maximally entangled configurations. 

We observe that our method identifies states minimizing the potential of multipartite entanglement in Eq.\eqref{eq:pi_7} which resulted globally more entangled than the states maximizing the number of bipartitions in which entanglement can be simultaneously maximized identified in \cite{Huber_2017}.

Internal energy and entropy are shown in Fig.\ \ref{fig:numerical-panel7}. The predicted bound for the number of MMES is subject to a systematic shift. 
The shift arises from the discrepancy between the $U(\beta \rightarrow +\infty)$ in Table~\ref{tab:mean_value&lowest_value U} and lowest value computed for the potential of multipartite entanglement, see Eq.~(\ref{eq:pi_7}). At the highest---yet finite---value of $\beta$, the internal energy remains significantly larger than the lowest value computed for the potential of multipartite entanglement~(\ref{eq:pi_7}), which indicates the presence of a residual population of less entangled (i.e.\ higher energy) states  in the computation of  internal energy and entropy. The discrepancy is the product of the computational cost and finite simulation time of the tempering protocols.
Nevertheless, the procedure introduced here provides a reliable leading-order estimate for the size of this class, namely, on the order of 2.5 million states. This figure should be  compared with the cardinality of the full set of hypergraph states, which is approximately $1.70 \times 10^{38}$.

\section{Discussion}\label{sec:disc}

To summarize the statistical behavior of the physical system under consideration, and consequently its entanglement properties, it is convenient to eliminate the control parameter $\beta$. This parameter has been useful in Figs.\ \ref{fig:comp_345}, \ref{fig:numerical-panel6} and \ref{fig:numerical-panel7}, as it is directly related to annealing protocols where temperature plays a central role, but it does not represent an intrinsic thermodynamic variable of the system.
The statistical properties of entanglement can instead be directly discussed in terms of the canonical quantities obtained from the numerical protocol \cite{Facchi_phase_trans_meta_2010}.

\begin{figure}[t]
    
     \includegraphics[width=0.48\textwidth]{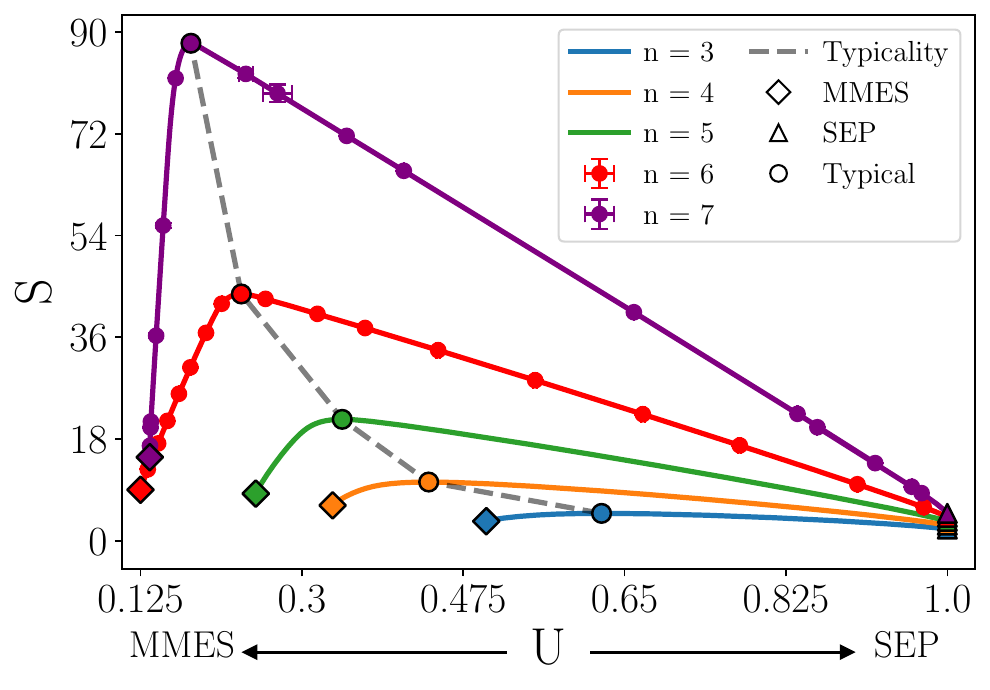}%
    \caption{\justifying Entropy $S$ as a function of the internal energy $U$ (average purity) for different number of qubits. For $n=3,4$, and $5$ qubits, the results are obtained from the complete enumeration over the hypergraph sets. For $n=6$ and $7$ qubits, all data points are generated using the numerical protocol based on annealing. The diamond-labeled MMES identifies the points along the curves corresponding to maximal multipartite entangled state configurations, the triangle-labeled SEP denote the points associated with separable configurations, while the circle-labeled ``Typical" indicates the points where typical states are reached. The dashed line connecting the points corresponding to typical states shows that $S\propto U^{-2}$, in agreement with the theoretical prediction from~\eqref{eq:mean_u} and~\eqref{eq:N_hyper}.}
    \label{fig:UvsS}
\end{figure}

For each system size, there exists a one-to-one correspondence between the internal energy and the entropy, allowing for a direct characterization of the entropy vs energy relation without explicit reference to temperature. In this framework, the thermodynamic behavior becomes particularly transparent when the entropy is plotted as a function of the internal energy, as shown in Fig.~\ref{fig:UvsS}. From a physical point of view, Fig.~\ref{fig:UvsS}. displays (the logarithm of) the number of states at a given value of average entanglement (purity).

This representation removes the inconvenience associated with the strongly scale-dependent range of $\beta$. Moreover, it enables a direct comparison of thermodynamic states across different system sizes, since the entropy grows only logarithmically with the cardinality $N_{\mathrm{hyper}}$ of the hypergraph set. The different classes of quantum states, which appear as plateaus in 
Figs.\ \ref{fig:comp_345}, \ref{fig:numerical-panel6} and \ref{fig:numerical-panel7}, are now mapped onto specific regions of the curve $S(U)$. In particular, MMES, typical states (uniform random ensemble) and separable states occupy distinct locations across the entropy-energy diagram, providing a unified thermodynamic characterization of the entanglement structure of the set.

\section{Conclusion}\label{sec:conclusion}

We have considered a statistical-mechanical framework for analyzing multipartite entanglement in quantum systems of qubits, restricted to the hypergraph state set. By mapping the average bipartite purity, used here as a measure of multipartite entanglement, onto an effective classical Hamiltonian defined over $2^n$ binary spin configurations, we established a thermodynamic description that captures the structure of the entanglement landscape through statistical observables such as internal energy and entropy. Within this formulation, the temperature acts as a tunable control parameter interpolating between the uniform ensemble of random hypergraph states at high temperature and the maximally multipartite entangled states (MMES) in the zero-temperature limit.

For small systems ($n \leq 5$), the hypergraph state set allows for an exhaustive enumeration of all microstates, providing exact thermodynamic quantities and enabling a direct validation of the theoretical framework. For larger systems ($n \geq 6$), where the exponential growth of the configuration space renders full enumeration impractical, we developed annealing-based numerical protocols, based on simulated annealing and parallel tempering algorithms, capable of efficiently sampling the relevant regions of the state space. These approaches allowed us to extensively explore the intricate structure of multipartite entanglement within the set of hypergraph states of qubit systems. Our results confirm that the thermodynamic analogy captures the essential features of multipartite entanglement within structured quantum ensembles. In particular, it reveals distinct fixed points associated with separable and maximally entangled configurations, thereby providing a statistical interpretation of entanglement complexity. The proposed framework therefore provides both a conceptual bridge between quantum information theory and classical statistical mechanics and a practical computational strategy for exploring entanglement in constrained many-body systems.

Looking forward, this approach opens several research directions. Extending the method to other structured ensembles, beyond the hypergraph state set, and investigating phase-like transitions in the entanglement landscape could yield further insight into the interplay among frustration, typicality, and complexity in multipartite quantum systems. The framework developed here thus establishes a foundation for studying the thermodynamic behavior of entanglement in large-scale quantum architectures, with potential implications for quantum simulation, quantum computation, and the statistical physics of complex quantum networks.

\section*{Acknowledgments}

We thank Chiara Macchiavello and Flavio Baccari for useful discussions and feedback on our work.
We acknowledge support from INFN through the project ``QUANTUM" and from the Italian funding within the ``Budget MUR - Dipartimenti di Eccellenza 2023–2027" - Quantum Sensing and Modelling for One-Health (QuaSiModO).
We acknowledge  support from Projects PN-RIC Q-SUD (Q-SUD B99H26000380007) and PN-RIC QUANTAS (SYNERGIA B99H26000410007).
PF acknowledges support from the Italian National Group of Mathematical Physics (GNFM-INdAM). GM acknowledges support from the University of Bari via the 2023-UNBACLE-
0244025 grant and from INFN through the project ``NPQCD".
We acknowledge computational resources provided by the University of Bari and the INFN cluster 
ReCaS~\cite{RECAS}. KZ acknowledges financial support from the European Union through ERC Advanced Grant TAtypic, Project No. 101142236.

\appendix
\let\clearpage\relax

\section{Annealing Methods for Hyper-graph states}
\label{sec:appA}

\begin{figure*}[t]
    \centering
 
    \subfloat[]{%
        \includegraphics[width=0.4\textwidth]{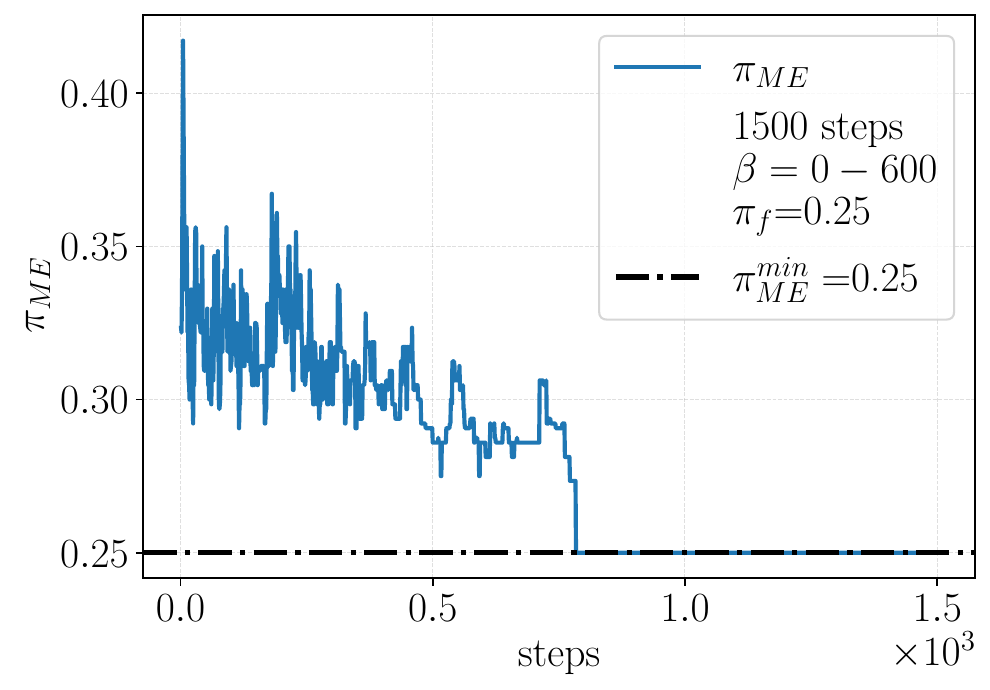}%
    }\hspace{0.01\textwidth}
    \subfloat[]{%
        \includegraphics[width=0.4\textwidth]{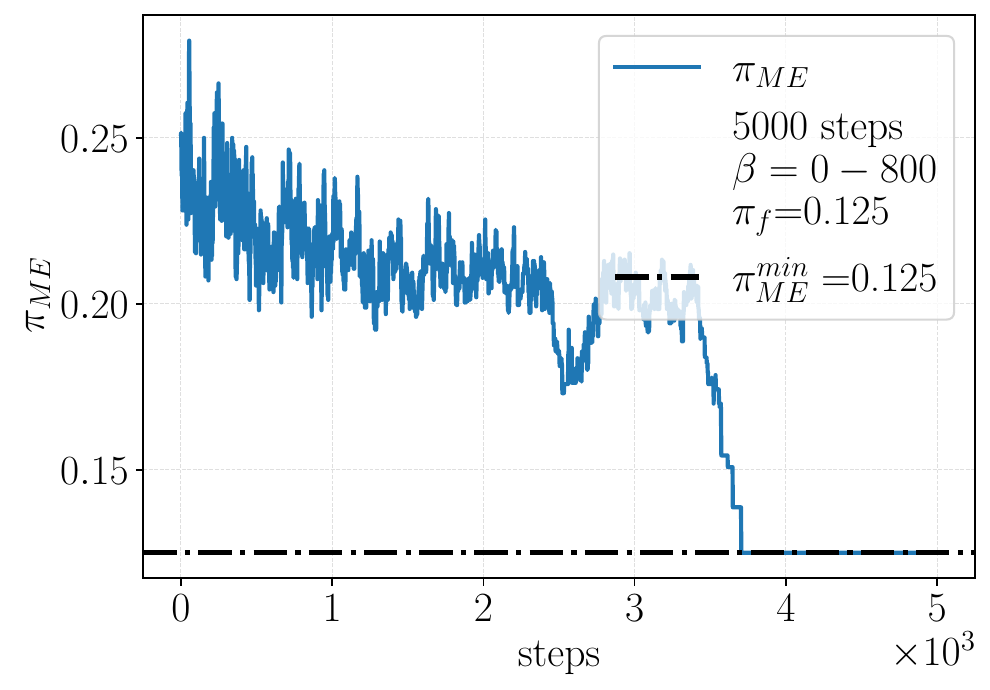}%
    }\hspace{0.01\textwidth}
    
    \caption{Optimization of $\pi_{\textrm{ME}}$ using the Annealing Method:
    (a) for 5 qubits, with $\beta$ increasing from  0 to 600 and 1500 annealing steps, and
    (b) for 6 qubits, with $\beta$ increasing from 0 to 800 and 5000 annealing steps.
     }
    \label{fig:annealing-plots}
\end{figure*}

For systems with $n \geq 6$ qubits, the exponential growth of the state space renders exact enumeration computationally intractable. To overcome this limitation and investigate the thermodynamic behavior of the hypergraph state set, we consider annealing strategies, exploiting the mapping between the hypergraph state ensemble and a classical spin model of $2^n$ binary variables. Within this framework, each state in the set is uniquely represented by a set of coefficients ${ s_k }$, each taking values $\pm 1$, which collectively determine the multipartite entanglement potential $\pi_{\textrm{ME}}$. This potential is interpreted as the Hamiltonian of the corresponding classical spin system, Eq.~(\ref{eq:hamiltonian}).

The annealing algorithm \cite{metropolis_1953, Annealing_1983, cerny_1985, Laarhoven_1987, Ingber_1993} seeks to minimize this cost function through an iterative process of single-spin flips. Proposed updates are accepted or rejected probabilistically according to the Metropolis criterion, ensuring a controlled exploration of the configuration space. The acceptance probability for a proposed move is given by:
\begin{equation}
x < e^{-\beta (E_2 - E_1)},
\end{equation}
where $x$ is a random number uniformly distributed in $[0,1]$, $\beta$ is the inverse temperature (with $k_B = 1$), and $E_1$, $E_2$ denote the energies before and after the spin flip, respectively.

This acceptance rule guarantees that spin flips leading to a lower energy ($E_2 < E_1$) are always accepted, while those increasing the energy are accepted with a probability that decreases exponentially with the energy gap and increments of $\beta$. At small values of $\beta$, the dynamics allow nearly unconstrained exploration of the configuration space, facilitating escape from local minima. As $\beta$ increases, the algorithm progressively favors energetically downhill moves, guiding the system toward low-energy configurations corresponding to highly entangled states.

An effective annealing schedule begins at a low inverse temperature $\beta_{\text{min}}$, typically   $\beta = 0$, where the system freely explores the configuration space, and gradually increases $\beta$ up to a maximum value $\beta_{\text{max}}$ over a predefined number of iterations or sweeps. The cooling schedule is chosen to balance exploration and exploitation, typically following a linear or exponential increase in $\beta$. At each temperature step, the system undergoes multiple Monte Carlo sweeps, ensuring adequate thermalization at each stage. 

The performance of the annealing optimization is illustrated in Fig.~\ref{fig:annealing-plots} for 5- and 6-qubit hypergraph state systems, demonstrating the method’s ability to approximate the entanglement landscape and thermodynamic observables even in high-dimensional settings where exact methods are computationally prohibitive. 

Parallel Tempering \cite{Tempering_1986, Geyer_1991, Marinari/Parisi_1992, Tempering_1996, Falcioni_1999, Grigera_2001, Earl_2005} shares most of its features with standard Simulated Annealing. The main difference lies in the introduction of multiple \emph{replicas} of the system, all evolving in parallel in order to minimize the chosen cost function.
In Parallel Tempering, on which we based our protocol, each replica evolves at a fixed temperature (i.e.\ fixed $\beta$), with different replicas assigned to different inverse temperatures. The strength of this approach stems from the possibility of periodically attempting swaps between replicas at neighboring temperatures. These swap moves are accepted with a probability that depends on both the energies of the replicas and the temperatures at which they evolve. The acceptance condition is given by
\begin{equation}
x < e^{-\Delta_{\beta} \Delta_E},
\end{equation}
where $x$ is a random number uniformly distributed in $[0,1]$, $\Delta_{\beta}$ denotes the difference in inverse temperatures between two neighboring replicas (with $k_B = 1$), and $\Delta_E$ is the difference between their energies prior to the swap attempt.
This exchange mechanism allows configurations generated at high temperatures—where barriers can be more easily overcome—to propagate toward lower temperatures, thereby enhancing the exploration of configuration space and mitigating trapping in local minima.

\section{Expectation value of $\pi_{\mathrm{ME}}$}
\label{sec:appC}

The energy at $\beta = 0$, denoted as $\textrm{U}(0)$, corresponding to the expectation value of $\pi_{\mathrm{ME}}$ over the complete ensemble of hypergraph states, can be analytically computed, as shown in \cite{Facchi_Lincei_2009},\cite{Trotta_2026}, leading to the following result:
\begin{equation}
\langle \pi_{\textrm{ME}} \rangle = \frac{ 2^{\lceil n/2 \rceil} + 2^{\lfloor n/2 \rfloor} - 1}{2^n}.
\label{eq:mean_u}
\end{equation}
In our annealing methods, despite extensive sampling, the numerical results still represent approximations of the true statistical expectation values, as the size of the hypergraph state sets becomes prohibitively large already with $n = 6$ qubits. Nevertheless, the very small relative errors reported in Table \ref{tab:mean_value&lowest_value U} demonstrate that our numerical estimates are in practice indistinguishable from the theoretical predictions, validating the accuracy of our annealing-based simulation strategies.

\renewcommand{\refname}{References}

\bibliography{References}

\end{document}